\documentclass[12pt]{iopart}
\usepackage{graphicx}

\begin{document}

\title[Early 
emission from Gamma-ray Bursts: from Gamma-ray to X-ray]
{Early multi-wavelength 
emission from Gamma-ray Bursts: from Gamma-ray to X-ray}

\def\swift{{\it Swift}}

\author{
P.T. O'Brien,
R. Willingale,
J.P. Osborne,
M.R. Goad
}
\address{Department of Physics \& Astronomy,
University of Leicester, Leicester LE1 7RH, UK}
 
\begin{abstract} The study of the early high-energy emission from both long
and short Gamma-ray bursts has been revolutionized by the \swift\
mission. The rapid response of \swift\ shows that the non-thermal
X-ray emission transitions smoothly from the prompt phase into a
decaying phase whatever the details of the light curve. The decay is
often categorized by a steep-to-shallow transition suggesting that the
prompt emission and the afterglow are two distinct emission
components.  In those GRBs with an initially steeply-decaying X-ray
light curve we are probably seeing off-axis emission due to
termination of intense central engine activity.  This phase is usually
followed, within the first hour, by a shallow decay, giving the
appearance of a late emission hump. The late emission hump can last
for up to a day, and hence, although faint, is energetically very
significant. The energy emitted during the late emission hump is very
likely due to the forward shock being constantly refreshed by either
late central engine activity or less relativistic material emitted
during the prompt phase. In other GRBs the early X-ray emission decays
gradually following the prompt emission with no evidence for early
temporal breaks, and in these bursts the emission may be dominated by
classical afterglow emission from the external shock as the
relativistic jet is slowed by interaction with the surrounding
circum-burst medium. At least half of the GRBs observed by \swift\
also show erratic X-ray flaring behaviour, usually within the first
few hours.  The properties of the X-ray flares suggest that they are
due to central engine activity. Overall, the observed wide variety of
early high-energy phenomena pose a major challenge to GRB models.

\end{abstract}

\pacs{98.70.Rz, 95.85.Nv, 95.85.Pw, 97.60.Lf}
\submitto{\NJP}
\maketitle

\section{Introduction}

Gamma-ray bursts (GRBs) are detected as bright, brief flashes of
gamma-rays which occur at some random location on the sky.  For a
short time, typically a few tens of seconds, the GRB is the brightest
single object in the gamma-ray sky, and more importantly is the
intrinsically brightest object in the Universe.  It is now generally
accepted that long-duration GRBs result from the death of a
rapidly-rotating massive star (a collapsar) while short-duration GRBs
arise from a merger of two compact objects, most likely two neutron
stars or a neutron star and a black hole (see \cite{pa98, mac99, mes2002,
zm2004} and references therein).  Either the collapsar or merger
 result in a black hole fed for a short time by an accretion
disk or torus. The accreting black hole can somehow power a
relativistic jet, presumably oriented along the rotation axis of the
black hole. The jet contains a relatively modest amount of baryonic
material moving at very high Lorentz factor -- the fireball.  Within
the jet the flow is not homogeneous, leading to internal shocks which
produce the initial, prompt gamma-rays that can be viewed if our
line-of-sight lies within the jet beam. As the jet moves out from the
progenitor, it also encounters circum-stellar and inter-stellar
material which results in classical afterglow emission produced by an
external shock (\cite{zm2004} and references therein).  We observe
some combination of these emission components and require as
continuous and lengthy an observation as possible in order to
disentangle them and hence test GRB models.

In this article we discuss the early gamma-ray and X-ray emission from
GRBs, concentrating on the observed temporal and spectral behaviour as
the GRB evolves over the first few hours.  Early GRB
observations have been revolutionized following the launch of the
\swift\ satellite on 20 November 2004 \cite{ge04}.  Although
gamma-ray emission can typically be detected by the Burst Alert
Telescope (BAT; \cite{ba05a}) on \swift\ for only a few tens of
seconds, the satellite can rapidly ($\sim 100$s) slew to point its
Ultraviolet and Optical Telescope (UVOT; \cite{ro05}) and X-ray
Telescope (XRT; \cite{bu05a}) at the GRB. The XRT permits observations
in the 0.3--10 keV band. This capability has ended what might be
termed the``X-ray dark ages'' for GRBs as previous missions rarely
obtained X-ray data in the period from a few minutes to a few hours.
It is this capability that we will exploit to describe the early
high-energy emission from GRBs.

In section 2 we provide an historical overview and outline why
\swift\ was built. The observational results are summarised in section
3, in which the emphasis is on the new phenomena revealed by \swift .
Conclusions are given in section 4.

\section{Historical overview}

Gamma-ray bursts were first announced as MeV events lasting between
0.1 and 30s, not from the Earth or Sun \cite{kle73}.  Their
discovery led to the inclusion of the BATSE instrument on the {\it
Compton Gamma-Ray Observatory}, which was operational between 1991 and
2000. See \cite {fis95} for a BATSE-era review.
BATSE acted as an all sky monitor over 20--600 keV, detecting
around 2,700 GRBs. The striking isotropy of these GRBs indicated an
origin either very close by or at cosmological distances, but the
BATSE GRB position uncertainties of 4 arc-minutes or larger prevented
identification of possible counterparts in other regions of the
electromagnetic spectrum. Even so, tantalising details did emerge from
the large BATSE GRB sample: the burst duration distribution was found
to be bimodal, with a population of short and spectral hard bursts
having durations of around 0.1--2s and a larger group of slightly
softer bursts with typical durations of 10--100s \cite{kou93}; and
the distribution of burst intensities was non-Euclidian, pointing to a
distance effect in the population.  

Over the BATSE bandpass, GRB spectra were shown to be non-thermal
and usually well fitted by a broken power law or Band function
\cite{band93, preece00}. A minority of the GRBs detected by BATSE were also
detected at higher energies by the EGRET or COMPTEL instruments on
{\it CGRO}. Aside from a few exceptions \cite{gon03}, the
extrapolation of the Band function fitted the very high energy spectra
well \cite{wink92, hanlon94}.  The burst
profiles, which were highly variable, defied classification.

It was not until the {\it Beppo-SAX} satellite (1996--2002) coded-mask hard
X-ray Wide Field Camera detected bursts, that a GRB was rapidly observed
with an imaging X-ray telescope.  The discovery of a fading X-ray
afterglow to GRB970228 \cite{cos97} ushered in a new era in which
positions sufficiently precise became available quickly enough for
ground-based follow up observations. Faint optical afterglows were
discovered, and it was quickly established that GRBs occurred at very
large distances (the first redshift was measured for GRB970508 at
$z=0.835$
\cite{met97}). 

With the distance scale known, the energetics of GRBs were thrown into
sharp relief. The observed fluences (the flux integrated over the
duration of the burst) and redshifts led to isotropic energies of
$\sim 10^{52} - 10^{54}$ erg, comparable to or larger than those of
supernovae. The very high gamma-ray luminosity leads to a compactness
problem caused by the high electron-positron pair production rate.
This can be solved by invoking a very high outflow velocity, $v$ with
Lorentz factor, $\Gamma_{jet} = (1-(v/c)^2))^{-0.5} \sim 100$ -- 300
\cite{baring93, fen93}.

A connection with supernovae was soon revealed by GRB980425/SN1998bw,
although this was an atypically nearby and low energy burst
\cite{gal1998, kul1998}. The supernova 
nevertheless showed very high velocities (tens of thousands of kilometers 
per second), and was given a new classification as a hypernova. 
The more typical GRB030329 confirmed the hypernova connection \cite{sta2003, 
hjo2003}, firmly establishing the collapsar model for long GRBs. 
Optical observations of GRB afterglow decays also showed a break to an
increased decay rate at around a few days. This was taken to be a sign
that the relativistic beaming of the slowing collapsar ejecta had
declined to the opening angle of the ejecta jet, $\theta_{jet}$, derived to
be $\sim 3$ -- 40
degrees, \cite {bfk2003}). 
This evidence for a jet rather than isotropic ejecta simultaneously reduced 
the energetics of the GRB to $\sim 10^{51}$ erg, 
and of course increased the implied GRB production rate. 
See \cite{zm2004} for an
excellent review of GRB knowledge before the launch of \swift . 

Prior to \swift , afterglows had only been securely detected for long
GRBs.  The expectation was that short GRBs might have a different
progenitor (e.g., a neutron star -- neutron star collision), however the
lack of any short GRB afterglows and hence precise positions for these
bursts had prevented the kind of progress made with the long
GRBs. \swift\ was created to provide accurate and prompt positions for
both short and long GRBs. The great difficulty of obtaining early
X-ray observations with existing satellites had led to very poor
knowledge of the afterglow behaviour before around 8 hours. \swift\ has
shown the richness of this interval, and has clarified the nature of
the short GRBs. There is much still to be understood, but \swift\ is
delivering a wonderfully rich scientific return.

\begin{figure}
\begin{center}
\includegraphics[angle=-90,width=.5\textheight]{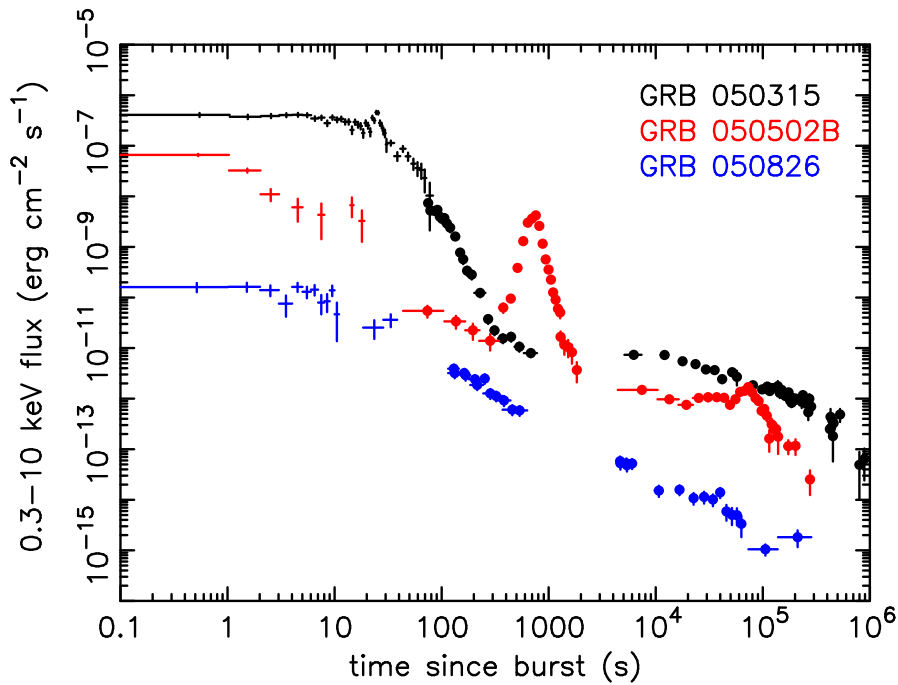}
\end{center}
\caption{
Example bursts showing the various behaviour patterns seen by \swift : a
steep-to-shallow transition (GRB050315, dark upper points); a large
X-ray flare (GRB050502B, middle light points); and a gradually
declining afterglow (GRB050826; lower points, divided by 100 for
clarity). BAT data are shown as crosses, XRT data as filled circles. 
}
\label{lcexample}
\end{figure}

\section{Observations in the \swift\ era}

Since launch \swift\ has detected an average of 2 GRBs per week. The
standard sequence of observations starts with detection by the BAT. The
on-board software then determines if it is safe to slew, and if so
commands the spacecraft to turn and point its narrow-field instruments at
the burst location. The slew typically takes 1--2 minutes. Thus, for the
longest duration bursts, XRT and UVOT observations can begin while the BAT
is still detecting the GRB. This capability of rapid, autonomous response
provides data which have revealed a wealth of phenomena in the X-ray
afterglow. \swift\ has detected a wide range of bursts including the
highest redshift GRB to date \cite{wat06} and the first localization of a
short burst \cite{ge05}.

We adopt the convention here 
of describing GRBs as long or short in terms of the 
timescale over which 90\% of the gamma-rays were detected --- the
T$_{90}$ parameter. GRBs with T$_{90}$ greater than or less than 2s
are denoted long or short bursts respectively \cite{kou93}.  The GRB
X-ray flux can be represented as a function of time and frequency
using a function $f_\nu\propto \nu^{-\beta} t^{-\alpha}$, where
$\beta$ is the spectral index and $\alpha$ is the temporal index. The
photon index $\Gamma$ is related to $\beta$ by $\Gamma = \beta + 1$.

The \swift\ data presented here were processed using the standard
analysis software.  The BAT data were processed using \swift\
software v2.0 as described in the BAT Ground Analysis Software Manual
\cite{kr04} and then light curves and spectra were extracted over
15--150 keV.  Power laws were fitted over the T$_{90}$ period to
provide spectral indices ($f_\nu\propto \nu^{-\beta_b}$). In most
cases a single power law provides a statistically acceptable fit
(i.e., reduced chi-squared, $\chi_{\nu}^{2} \le 1$), although on
occasion a cutoff power law provides a better fit.  Similar power law
fits were used to parameterise the XRT spectra ($f_\nu\propto
\nu^{-\beta_x}$), over 0.3--10 keV. For many GRBs intrinsic absorption
in addition to the Galactic column is required to provide a good
fit. The required intrinsic column is in the range 2--35 $\times 10^{22}$ 
cm$^{2}$ 
\cite{ca06, ob06}.

\begin{figure}
\begin{center}
  \includegraphics[angle=-90,width=.6\textheight]{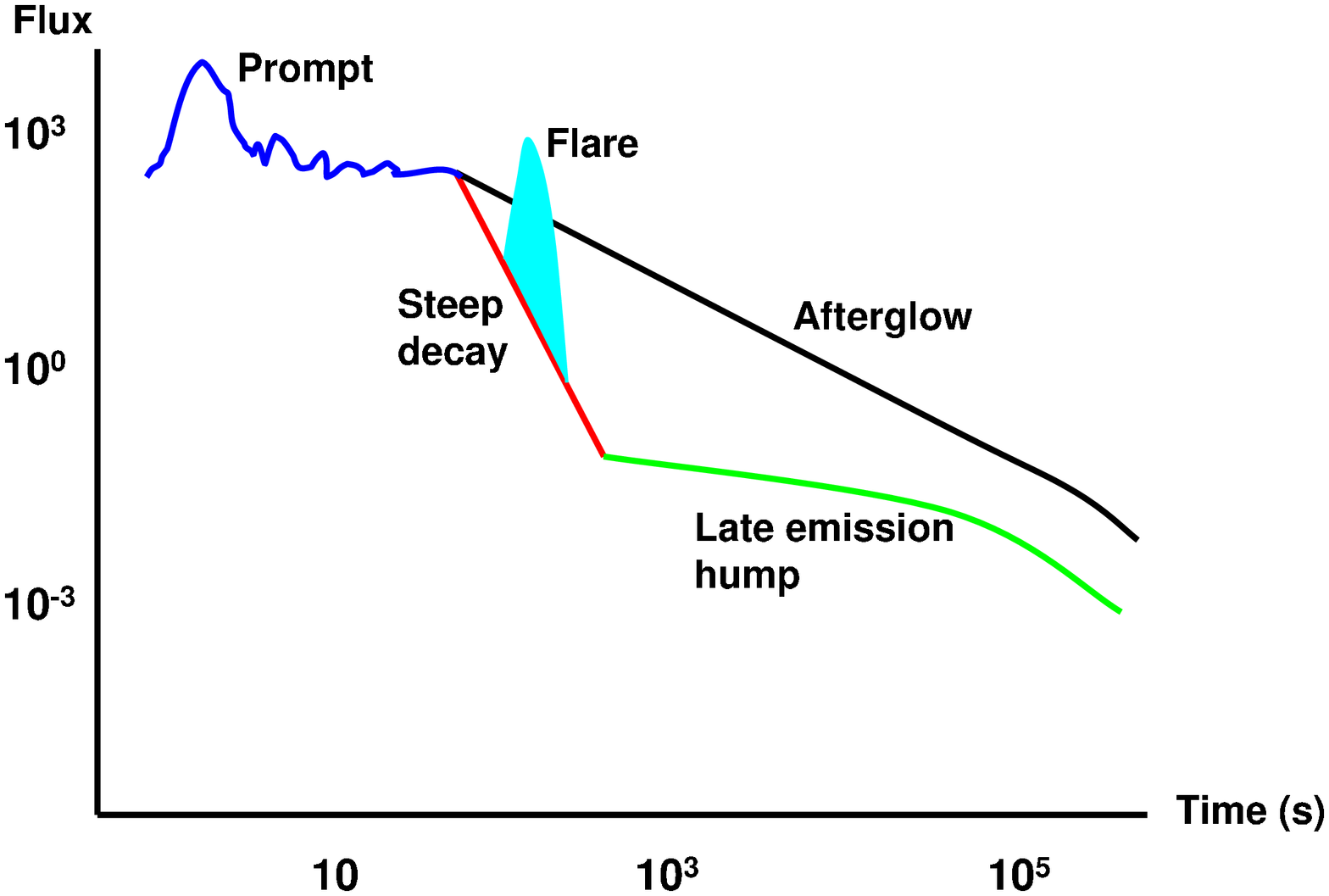}
\end{center}
  \caption{A schematic view of the early GRB X-ray light
  curve. Following the prompt emission, which typically lasts a few
  10s of seconds, the decay tends to follow one of two paths: (i)
  a steep decay (flux $\propto t^{-3}$), during which the flux can fall
  by 3 or more orders of magnitude, followed by a shallower, late
  emission hump ($\propto t^{-0.5}$) starting at $\sim 10^3$s; or (ii)
  a gradual decay ($\propto t^{-1}$). Either decay path can end with a
  break at $>10^4$s to a steeper decay. X-ray flares can occur during
  either decay path, most prominently during the first hour. See text
  for details.}
\label{schematic} 
\end{figure}

Analysis of a large GRB sample \cite{ob06} shows that the XRT spectra
of GRBs usually require a softer power law than the BAT spectra
(i.e. $\beta_x > \beta_b$).  To form unabsorbed, 0.3--10 keV flux
light curves for each GRB, we therefore (a) converted the XRT count
rates into unabsorbed fluxes using the XRT power law spectral model
and (b) converted the BAT count rates into unabsorbed fluxes by
extrapolating the BAT data to the XRT band using a power law spectral
model with an absorbing column derived from the XRT data and a
spectral index which is the mean of the XRT and best-fit BAT spectral
indices.  In those GRBs which have high signal-to-noise ratio data,
more complex spectral evolution can be seen, but the above procedure
has been applied for consistency for all bursts discussed in this
paper.

The initial \swift\ results appeared contradictory; some long-duration
bursts displayed a rapid decline in the first hour, with temporal
decay indices, $\alpha \ge 3$ \cite{ta05, va06}, while in others the
early X-ray flux declined more gradually with $\alpha \sim 1$
\cite{ca05}.  Several of those with a steep decline also displayed a
shallower decay starting within an hour and lasting up to a day
from trigger.  A large fraction of GRBs also have X-ray ``flares''
during the first few hours superimposed on the declining light curves.
Examples of the various observational phenomena are shown in
Fig.~\ref{lcexample}. Here, GRB050315 shows a steep decline followed by a
long shallower decay before breaking again at late times; GRB050502B
displays a large X-ray flare, while GRB050826 shows a gradual decline after
the prompt emission.  As more bursts have been observed a pattern has
emerged which is summarised schematically in Fig.~\ref{schematic} (see
also \cite{nou06, zhang06, ob06}).  Each of these phenomena are
discussed below, but the overall behaviour is as follows:

\begin{enumerate}

\item The ``prompt'' emission is that emitted
directly during the burst. With \swift\ this emission is seen by the
BAT but can also be detected by the XRT if the burst is long enough to
last until the completion of the first slew to target. Most bursts
observed by \swift\ typically have a 15--150 keV spectral index of $\beta_b =
0$ -- 2 during the prompt phase.

\item The prompt phase is followed by a power law decline phase. The first
temporal index, $\alpha_1$, during this phase can be very large (up to
$\approx 5$) and in most GRBs $\alpha_1 > 2$, but in a significant
minority, perhaps 20--30\%, a gradual decline is observed with $0.5
\le \alpha_1 \le 1.5$. The wide range in temporal index 
suggests several emission processes may be involved.  The spectral
index in the 0.3--10 keV band during this phase is usually in the
range 0.5--2.5, although occasionally larger values are seen.

\item For those bursts which initially decline steeply, the decay breaks
to a shallower rate, typically within the first hour, such that $0.5\le
\alpha_2 \le 1.5$. This ``late emission hump'', can last for up to $\sim
10^{5}$s before breaking to a steeper decay. The late emission hump
appears to have a harder spectrum on average than the steep decline
phase (section 3.3).  It can have a fluence equal to that of the
prompt phase \cite{ob06}, so although faint in observed flux, this
phase is energetically very significant.

\item For those bursts which initially decline gradually the temporal and
spectral indices are broadly consistent with a ``classical afterglow''
interpretation, in which the X-ray emission comes from the external
shock.  In these GRBs the late emission hump is usually not seen. This
does not mean that the late emission hump is absent as it may be hidden by the
classical afterglow component.

\item Limited statistics make quantifying later phases difficult, but both
the initially steeply declining bursts and those that decline more gradually
can show a late temporal break (typically at $10^4$ -- $10^5$s) to a
steeper decay. These late breaks are not seen in all GRBs --- some decay
continuously beyond $10^6$s until they fade below the \swift\ XRT
detection limit. There is usually no evidence for spectral changes during
late temporal breaks, which can be represented either as a series of
temporal breaks using multiple broken power laws or a smoothly curving
decay (e.g. \cite{va06}).

\item X-ray flares are seen in the first few hours for around half of the
GRBs observed by \swift , and occur in GRBs which decline rapidly or
gradually. The majority of these flares are only detected in the XRT but in
some bright, long bursts flares are observed simultaneously with the BAT.
Strong spectral evolution can be observed in some cases. Most of the X-ray
flares are energetically small, but a few are very powerful   
\cite{bu05b, os06, romano06} with a fluence comparable to
that of the prompt phase. Late flares are also occasionally seen.

\item The X-ray light curves for short bursts have been less well studied
by \swift\ as they are fainter (on average) and \swift\ has detected
fewer examples of short bursts. To date, the short burst light curves
display a range in phenomena remarkably similar to those seen in the
long bursts, including either rapid or gradual decay, flares and a
late emission hump \cite{ba05b, sod06}.

\end{enumerate}

The behavioural pattern of prompt emission followed by a steep X-ray
decay and then a shallow decay has been characterised as the
``canonical GRB light curve'' \cite{nou06}.  But, while this pattern
is seen in a majority of GRBs, as outlined above, it is not observed
in all.  To understand the various phases we need to consider each of
them in turn.

\begin{figure}
\begin{center}
  \includegraphics[angle=0,width=.6\textheight]{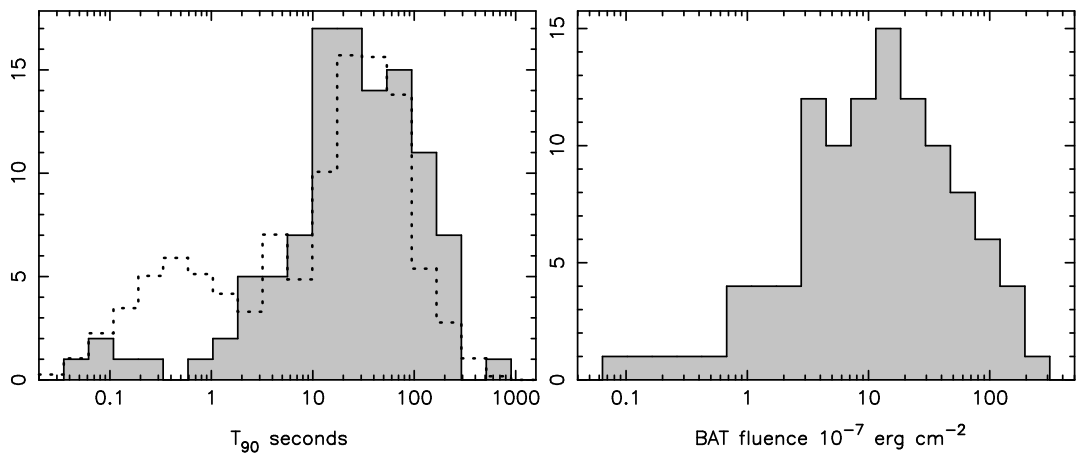}
\end{center}
  \caption{ The distribution of burst durations T$_{90}$ (left panel)
  and 15-150 keV fluence values (right panel) for GRBs detected by the
  \swift\ BAT. The dashed line in the left-hand panel is the T$_{90}$
  distribution for BATSE normalized to the number of \swift\ bursts.}
\label{bat} 
\end{figure}

\subsection{The prompt phase}

\begin{figure}
\begin{center}
  \includegraphics[angle=-90,width=.5\textheight]{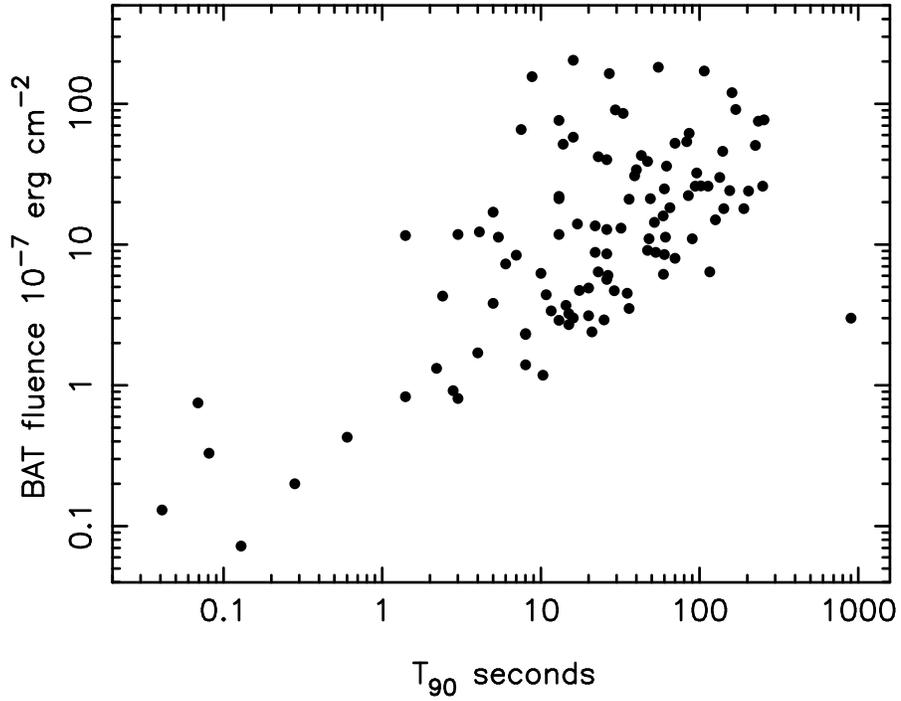}
\end{center}
  \caption{ Correlation of the 15--150 keV fluence with T$_{90}$ for
  GRBs detected by the \swift\ BAT.  The very long burst at far right
  is GRB 060123. The sensitivity of the BAT limits the detection
  of long faint GRBs.}
\label{batfluence} 
\end{figure}

BAT has detected and located on-board the prompt
emission of GRBs at a rate of approximately 100 yr$^{-1}$.  In terms
of duration, BAT GRBs span the same range as those detected by the
BATSE instrument, as illustrated in Fig.~\ref{bat}. The BAT data are for
those GRBs with values of T$_{90}$, 15--150 keV fluence and spectral
index available from the data table on the \swift\ web site
\cite{swift} as of mid-February 2006. 
While the parameters given in the data table are preliminary, the
shape of the distribution does not change significantly if data from
the forthcoming \swift\ catalog are used (T. Sakamoto, private
communication).  The BATSE data plotted in Fig.~\ref{bat} are those
from the revised 4B catalog \cite{paciesas99}.  Comparing the BAT and
BATSE distributions is difficult due to their different
energy-dependent sensitivities and trigger software \cite{band06}, but
as for BATSE, most \swift\ GRBs have durations of 10 to 100s and have
a 15--150 keV fluence within a factor of 10 of $2\times10^{-6}$ erg
cm$^{-2}$.  Fig.~\ref{batfluence} shows how BAT fluence correlates
with T$_{90}$, illustrating that the shortest GRBs have much less
fluence than the longest.

\begin{figure}
\begin{center}
  \includegraphics[angle=0,width=.6\textheight]{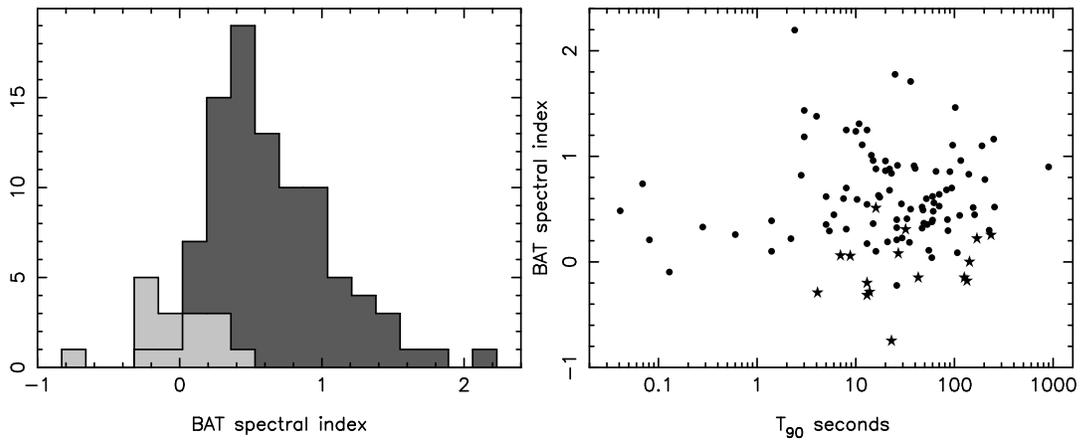}
\end{center}
  \caption{ The distribution of BAT spectral indices for
  GRBs. Left-panel: The spectra were fitted using either a single
  power law or a cutoff power law and are shown as the dark or light
  grey histograms respectively. The cutoff power law model was used
  when that improved the fit at $>99$\% confidence. Right-panel:
  Correlation of the BAT spectral index with T$_{90}$.  Spectra fitted
  using either a single power law or a cutoff power law are shown as filled
  dots or stars respectively. }
\label{batphoton} 
\end{figure}

The spectral indices derived from spectral fits
to the 15--150 keV BAT data are shown in Fig.~\ref{batphoton}.  Due to
differences between the BAT and BATSE detector energy bands and how
BAT uses rate triggers plus image-accumulation to find a point source,
the BAT is more sensitive than BATSE to long, soft bursts and detects
relatively fewer short, hard bursts than might be expected despite its
greater sensitivity to short triggers
\cite{band06}. Thus, the large majority of all BAT detected bursts
lie in the long, soft category, and include amongst them the highest
redshift bursts yet detected.  There is some indication from
Fig.~\ref{batphoton} that shorter bursts are spectrally harder, as
previously noted for BATSE bursts \cite{kou93}, but a much larger
sample is required to confirm this trend.

\subsection{The early decay phase}

One of the most surprising results from \swift\ has been the rapid decay
observed in many bursts starting, typically, within a few minutes of the
trigger. The steep decay rates, $t^{-\alpha}$ with $\alpha = 2-5$, are
significantly larger than those routinely observed in the optical or X-ray
for GRBs discovered pre-\swift , although it must be remembered that those
observations were usually at half a day or more post-trigger.

The early rapidly fading X-ray emission could have a variety of
possible explanations
\cite{zhang06}, including
high-latitude emission from the fading burst \cite{ku00}, the
interaction of the jet with the surroundings --- the classical
afterglow emission produced by an external shock \cite{me97}, or
thermal emission from a photosphere around the outflow \cite{me00} or
from a hot cocoon associated with the jet \cite{me01}.

For almost all of the GRBs observed by \swift\ the X-ray light curve
derived from the BAT data joins smoothly to that from the XRT.
If the BAT
and the XRT are initially detecting the prompt emission from the jet, when
this emission stops (for example the end of internal shocks)  we would
continue to observe photons coming from regions of the jet which are off the
line of sight --- the ``curvature effect'' or ``high-latitude emission''
\cite{ku00, tag05, nou06, zhang06, pan06, ob06}.
For such a model emission at angles $\theta$ from the line of
sight which are in excess of $\theta = \Gamma_{jet}^{-1}$ will start to
dominate the observed emission.
If the jet has uniform surface brightness, the observed X-ray
flux will fall as $t^{-\beta - 2}$ where the spectrum is $\propto
\nu^{-\beta}$. Thus this model predicts a relation such that $\alpha -
\beta =2$ for the early, rapidly declining part of the temporal decay. It
is possible to get a shallower decay if viewing a structured-jet off-axis
\cite{dyks06} although the general trend is similar to the standard
high-latitude model.  When considering high latitude emission, the
zero-time used to calculate the decay index need not correspond to the
trigger time if the light curve is dominated by a later event, such as a
large flare.

The possible contribution of standard afterglow emission as the jet
interacts with its surroundings complicates the comparison between
models and observations. Indeed as afterglow emission can begin within
minutes of the burst we are likely to be observing a mixture of
emission components, each contributing to the observed temporal and
spectral indices.  To disentangle the relative contribution of
emission from the central engine and that due to the
afterglow, O'Brien et al. \cite{ob06} systematically analysed the
temporal and spectral properties of a large GRB sample combining data
from the BAT and XRT. The sample comprised 40 GRBs detected by \swift\
prior to 2005 October 1 for which \swift\ slewed to point its
narrow-field instruments within 10 minutes of the burst trigger
time. Of the 40 GRBs, 38 are long bursts.

\begin{figure}
\begin{center}
\includegraphics[angle=-90,width=.5\textheight]{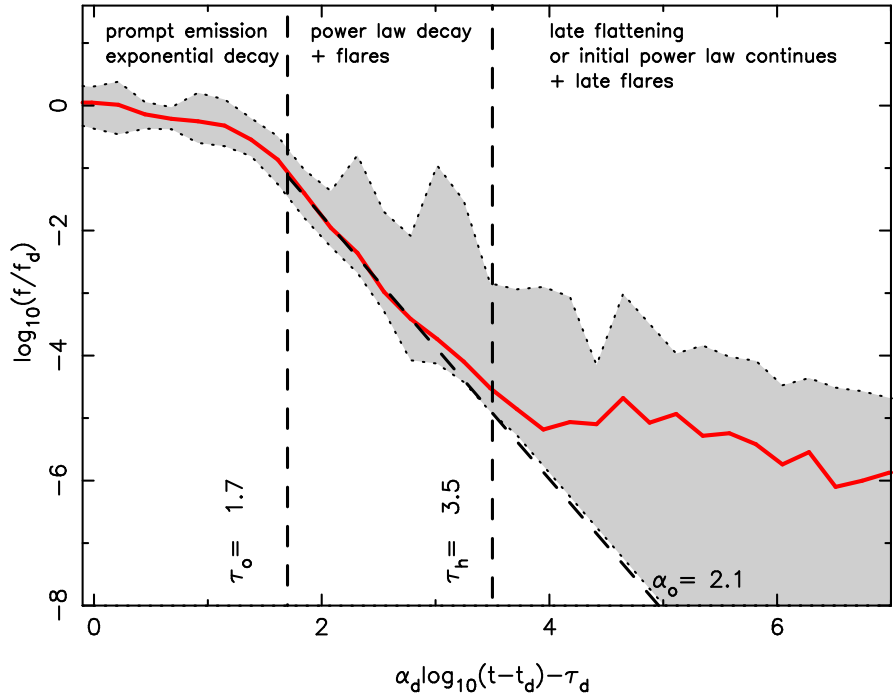}
\end{center}
  \caption{The composite X-ray light curve for 40 GRBs in \cite{ob06}
  for which there are BAT and early XRT data. The axis are normalized
  flux and transformed time units.  The average decay curve is shown
  as a solid curve and is well fit by an exponential for $\tau_{0} \le
  1.7$. It then relaxes to a power law with index $\alpha_{0}=2.1$,
  shown as a dashed line.  The shaded area bordered by dotted lines
  indicates the range of the individual flux values as a function of
  time. In this transformed space, those GRBs which show a gradual
  power law decline follow the average decay curve until close to
  $\tau_{h}$ and then continue to decline as power laws. The light
  curves for the majority of GRBs flatten above $\tau_{h}$.  About
  half of the GRBs exhibit sporadic flaring during the initial decay
  and/or the late period.  }
\label{lc}
\end{figure}

In order to compare light curves for GRBs with different power law
decay indices, O'Brien et al. \cite{ob06} developed a procedure to fit
light curves assuming there is a common intrinsic form to the early
X-ray light curve. An average X-ray decay curve expressed by log(time)
as a function of log(flux), $\tau(F)$, and log(flux) as a function of
log(time), $F(\tau)$, was derived by taking the sum of scaled versions
of each of the individual light curves, $f_{i}(t_{i})$, where $t_{i}$
is approximately the time since the largest/latest peak in the BAT
light curve.  The data points were transformed to normalised
log(flux), $F_{i}=\log _{10}(f_{i}/f_{d})$, and log(time) delay
values, $\tau_{i}=\alpha_{d}
\log_{10}(t_{i}-t_{d})-\tau_{d}$.  Four decay parameters (suffix {\it
d}) specify the transformation for each GRB: $f_{d}$, the mean prompt
flux; $t_{d}$, the start of the decay; $\tau_{d}$, a time scaling; and
$\alpha_{d}$, a stretching or compression of time.  The best fit
$f_{d}$, $t_{d}$, $\alpha_{d}$ and $\tau_{d}$ for each GRB were found
using a least squares iteration procedure, excluding bright
flares. The resultant composite light curve for the entire sample is
shown in Fig.~\ref{lc}.

\begin{figure}
\begin{center}
  \includegraphics[angle=-90,width=.5\textheight]{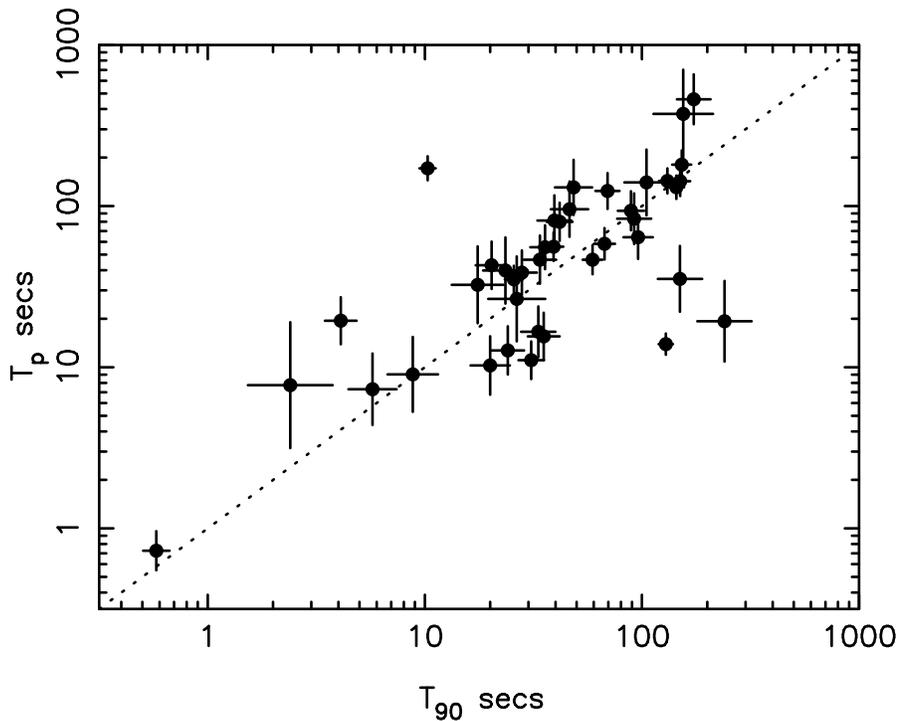}
\end{center}
  \caption{ Correlation of the duration of the prompt emission
  T$_{p}$, derived from fitting to the average 0.3--10 keV decay curve
  derived from the combined BAT+XRT data, with T$_{90}$, derived from
  the BAT 15--150 keV data.
 }
\label{tp} 
\end{figure}

Under the transformation all the light curves conform to an approximately
universal behaviour with an initial exponential decline $\propto
\exp(-t/t_{c})$ followed by a power law decay $\propto t^{-\alpha_{0}}$.
The transition between the two decay phases occurs when the
exponential and power law functions and their first derivatives are
equal, and is given for the average decay curve by
$t_{0}=t_{c}\alpha_{0}$ ($\tau_0 = 1.7$). Adopting this transition,
for each GRB we define the division between the prompt and power law
decay phases to be $\tau_{0}$, corresponding to a prompt time
T$_{p}=10^{(\tau_{0}+\tau_{d})/\alpha_{d}}$ seconds.  This prompt time
definition provides us with an alternative estimate of the duration of
the prompt phase for each burst which depends on the physical shape of
the BAT+XRT light curve rather than the sensitivity of the BAT.
As shown in Fig.~\ref{tp}, T$_{p}$ is comparable to
T$_{90}$ for many bursts, but it can be considerably shorter or
longer.

\begin{figure}
\begin{center}
  \includegraphics[angle=-90,width=.5\textheight]{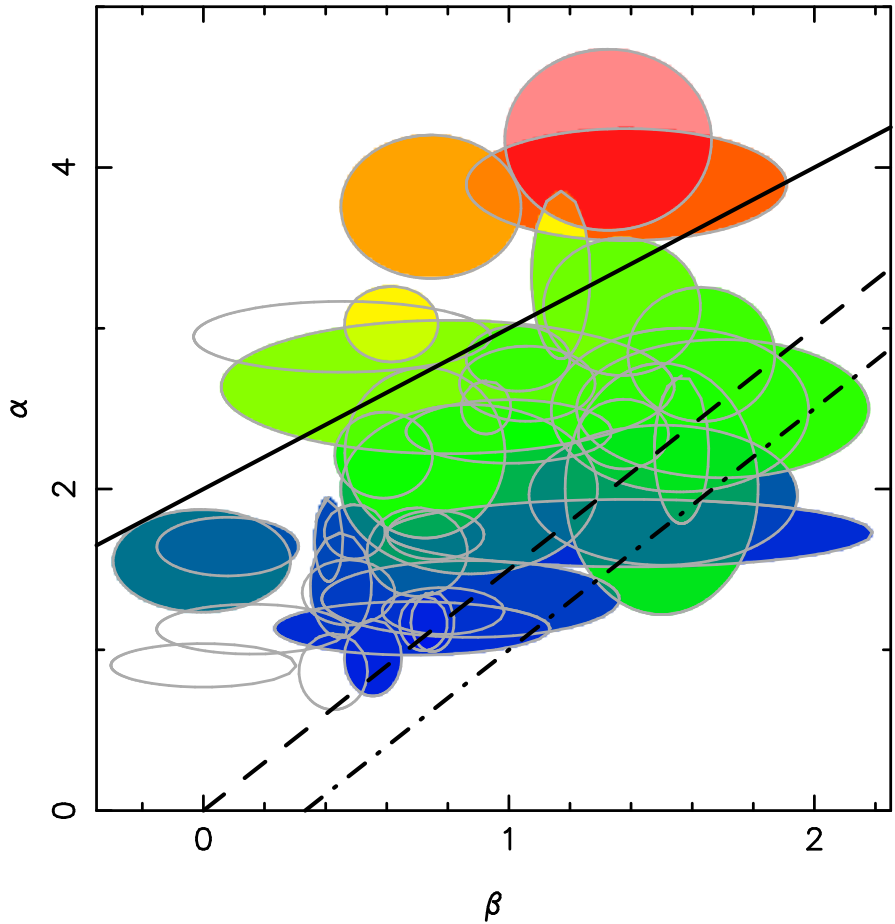}
\end{center}
\caption{
Correlation of the decay index, $\alpha$, with the spectral index,
$\beta$, for the 40 GRBs contributing to Fig.~\ref{lc}, where $\beta$
is the average of the spectral indices from the BAT and XRT.  Each GRB
is plotted as an ellipse representing the $90\%$ confidence region.
Blue indicates no late emission hump while red indicates a strong late
emission hump (see \cite{ob06} for details). Shades of green indicate
areas between these extremes. Open ellipses are GRBs for which there
are no late time data.  The solid line is the predicted relationship for
high-latitude emission. The dashed and dot-dashed lines are the
predictions for an afterglow model of a jet expanding into a constant
density medium before a jet break, observing in an energy band below
and above the cooling break respectively. Almost all of the GRBs lie
between the model predictions suggesting most objects have a
contribution from several emission components.  }
\label{corr}
\end{figure}

The average decay curve relaxes into a power law with a decay index
$\alpha_{0}=2.1$, found by linear regression on the average decay
curve for $\tau_{0}<\tau<3.0$. This power law fit is shown as a dashed
line in Fig.~\ref{lc}.  The fitting procedure results in those GRBs
which follow a fairly continuous decay lying close to the power
law. At $\tau \sim 3$ the average decay curve starts to rise above the
power law decay in the majority of bursts. This is the start of the
late emission hump, which we define to start at $\tau_h = 3.5$.

The initial temporal decay index for individual GRBs can be calculated
by multiplying $\alpha_{0}$ by the best fit $\alpha_{d}$.  GRBs with
$\alpha_{d}>1$ have decays steeper than average and those with
$\alpha_{d}<1$ shallower.  The resultant $\alpha =
\alpha_{0}\alpha_{d}$ are based on all the available data from both
the BAT and XRT and are expected to be a more robust estimate of the
initial power law decay rate than fitting a power law to a short
section of light curve.

The values of $\alpha$ and $\beta$ can be used to test the high
latitude and afterglow models, where $\beta$ is taken as the average
of the BAT and XRT spectral indices. The correlation between these
quantities is shown in Fig.~\ref{corr}.  In principle, the
relationship between the temporal decay index and spectral index has
two components such that $\alpha=\alpha_{\nu}\beta+\alpha_{f}$. The
coefficient $\alpha_{\nu}$ arises from the redshift of the peak of the
spectral distribution of the synchrotron emission as a function of
time and $\alpha_{f}$ arises from the temporal decay in the peak flux
value of the same spectral distribution.  The solid line in
Fig.~\ref{corr} shows the expected relationship for the high latitude
model with $\alpha_{\nu}=1$ and $\alpha_{f}=2$. The dashed line shows
the relationship expected for an afterglow model of a jet expanding
into a constant density medium observed at a frequency below the
cooling break ($\nu_x < \nu_c$) and before a jet break, with
$\alpha_{\nu}=3/2$ and $\alpha_{f}=0$
\cite{sa98}.  If $\nu_x > \nu_c$ then $\alpha_{\nu}$ is unchanged and
$\alpha_f = -0.5$. This is plotted as a dot-dashed line on
Fig.~\ref{corr}. All of the GRBs lie on or above these afterglow
lines. 
Very similar conclusions are reached if a wind afterglow model is adopted.

It is clear from Fig.~\ref{corr} that the decay and spectral indices
correlate well with the strength of the late emission hump. The bursts with
the most significant humps do not have large X-ray flares but they do
have steep decays and straddle the high latitude line \cite{ob06}.
The bursts with weaker humps lie below the high latitude line reaching
down to the afterglow lines. The majority of GRBs lie below the high
latitude prediction. For these it is likely that we are seeing a
combination of high latitude prompt emission and conventional,
pre-jet-break afterglow.

We note that those GRBs which decay more gradually are more likely to
have an early optical detection.  Using the initial \swift\ UVOT
V-band exposure to quantify the early optical brightness, for the GRBs
in our sample with UVOT observations in the first 10 minutes, those
with $\alpha < 2$ are four times more likely to have been optically
detected.

\subsection{The late emission hump}

Both the rapid-decay and classical afterglow models have difficulties
explaining the late emission hump.  Using the light-curve fitting
procedure described above, for $\tau > 3.5$, the maximum fluence of
the late emission hump is commensurate with the prompt fluence
\cite{ob06}, suggestive of some kind of equipartition in energy
between the emission phases.  A number of models have been proposed to
explain the late emission hump. It may be due to forward shock
emission, which is refreshed with energy either due to continued
emission from the central engine or because the ejecta has a range in
initial Lorentz factor \cite{re98, sari00, zhang01, nou06, zhang06,
granot06}.  As the injection process adds energy the decay does not
simply resume the previous decay curve following the shallow phase but
rather shows a step (Figs \ref{lcexample} and
\ref{schematic}).

\begin{figure}
\begin{center}
  \includegraphics[angle=0,width=.4\textheight]{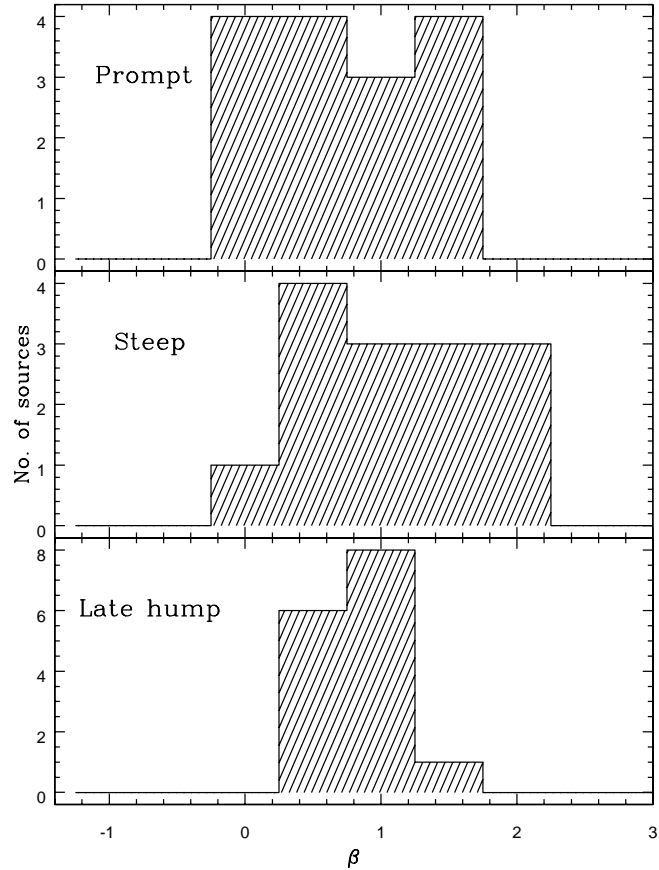}
\end{center}
  \caption{The observed distribution in spectral indices for the prompt
  phase (upper), steep decay phase (middle) and late emission hump
  (lower).  The spectral indices for the late emission hump
  are the most tightly clustered.
}
\label{photonhist}
\end{figure}

We can use the spectral characteristics to test possible relationships
between emission phases. Spectral index distributions for those
GRBs with a steep decay phase taken from \cite{ob06}, plus a few
others, are shown in Fig.~\ref{photonhist}. The prompt (BAT) spectra
have a mean spectral index of $0.61\pm0.02$ and standard deviation
$\sigma=0.59$.  The steep decay phase has a steeper mean spectral
index of $1.12\pm0.02$ and $\sigma=0.6$, while during the late
emission hump the mean spectral index is $0.86\pm0.03$, and
$\sigma=0.36$.  Interestingly, the late emission hump exhibits a far
narrower range in spectral shape than the earlier phases. The
narrowness of the spectral index distribution during this phase argues
in favour of a universal energy generation mechanism for this segment
of the light-curve.

The late emission hump spectra are, on average, harder than the
steep-decline spectra, supporting the concept of late-time shock
refreshing.  These two spectral indices are uncorrelated ($r=0.19$,
$P=0.49$). There is a much stronger correlation between the
difference in the spectral indices (steep-late hump) and the
steep-decline spectral index ($r=0.85$, $P=3\times 10^{-5}$).  There
is no correlation between the late emission hump spectral index and
the prompt spectral index ($r=0.28$, $P=0.3$). 

\subsection{Late evolution and jet breaks}

It has been previously shown that GRBs can show a
wavelength-independent late break in their optical and infrared light
curves \cite{rhoads97, sari99}. If this break is associated with the
jet slowing down, such that $\theta_{jet}$ becomes larger than
$\Gamma_{jet}^{-1}$ (the jet also starts to expand laterally), it can
be used to estimate the jet opening angle and hence the actual emitted
energy. The derived $\theta_{jet}$
imply a tightly clustered intrinsic, beaming-corrected, luminosity of
$\sim 10^{51}$ erg, which, if confirmed over a wider redshift range,
could allow the use of GRBs as standard candles \cite{frail01}.

A significant number of the GRBs found by \swift\ are at high
redshift. The mean redshift of the \swift\ sample is $<z> = 2.6$,
more than twice the
mean redshift $<z> = 1.2$ pre-\swift . This allows for a test of the
idea of GRBs as standard candles, but also poses a challenge for \swift .  
The late temporal breaks previously observed in the optical and
infrared occur at a few days, and will appear later, on average, for
\swift\ due to the increased time dilation. By this time the X-ray
count rates can be down by around five orders of magnitude, or more,
from peak. In the spectroscopy sample discussed above, only 9 GRBs
have sufficient counts to derive a spectral index after the end of the
late emission hump. Of these, 3 show spectral steeping, 5 shown no
evidence for spectral variation, and the remaining GRB is
inconclusive.  

The late temporal decay slope and the usual absence of clear spectral
variability suggests that this segment is associated with the normal
afterglow phase seen in pre-\swift\ bursts at those epochs. Currently
it is unclear if any jet breaks have been detected in long bursts
using data from \swift\ (although see \cite{blustin05}).  The
discovery of the late emission hump further complicates detection, as
the end of that phase could be mistaken for a jet break.  Sato et
al. \cite{sato06} analysed
\swift\ data for three GRBs with extended light curves and known
redshifts. They show that the bursts do not show an achromatic break
at the times expected, derived from an empirical relationship between
the peak in the energy spectrum of the prompt emission and the
isotropic luminosity \cite{ghirlanda04}. If confirmed for a larger
sample, this would indicate that the jet opening angle has a wider
dispersion than previously thought and hence GRBs have a wider range
in intrinsic luminosity.

\begin{figure}
\begin{center}

\includegraphics[angle=-90,width=.55\textheight]{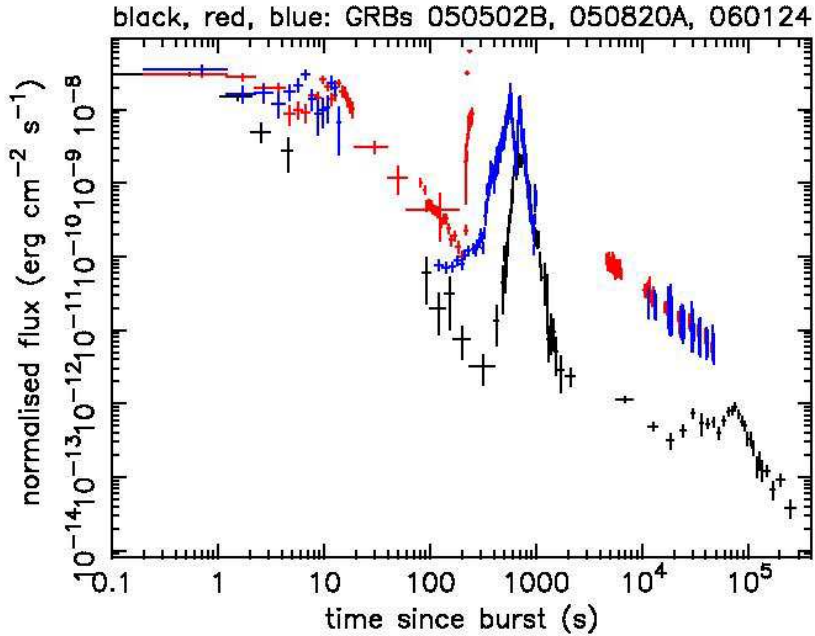}
\end{center}
  \caption{ The 0.3--10 keV light curves for GRB050502B (black),
  GRB060124 (blue) and GRB050820A (red) --- three GRBs with large
  X-ray flares. The fluxes have been normalised to the start of the
  prompt phase using the flux scale for GRB050820A. Observations of
  GRB050820A were interrupted by the passage of \swift\ through the
  South Atlantic Anomaly.  }
\label{bigflare}
\end{figure}

\subsection{X-ray flares}

The standard model for GRB afterglows, a spherical blastwave expanding
into a uniform density ambient medium, predicts smooth afterglow
light-curves. At least half of the GRBs observed by \swift\ show X-ray
flares \cite{ob06}. Suggested models for the origin of flares
include density fluctuations in the surrounding medium into which the
blastwave expands, structured jets, reverse shocks, refreshed shocks
and late-time central engine activity. These models predict relative
fluctuation amplitudes and timescales which can be used to rule out
some scenarios for the origin of the flares.

Three of the strongest X-ray flares, observed in GRB050502B,
GRB050820A and GRB060124, are shown in Fig.~\ref{bigflare}
\cite{bu05b, os06, romano06}).  These flares are quite late, and
beyond the burst duration measured by T$_{90}$. It can be argued that
the initial flux is a ``pre-cursor'', in which case the ``flare'' is
actually the burst. This is a distinct possibility, although we note
that without a flare all three light curves would still have been
classified as that of a GRB. This illustrates the uncertain definition
of pre-cursors, bursts and flares.
 
\begin{figure}
\begin{center}
  \includegraphics[angle=-90,width=.5\textheight]{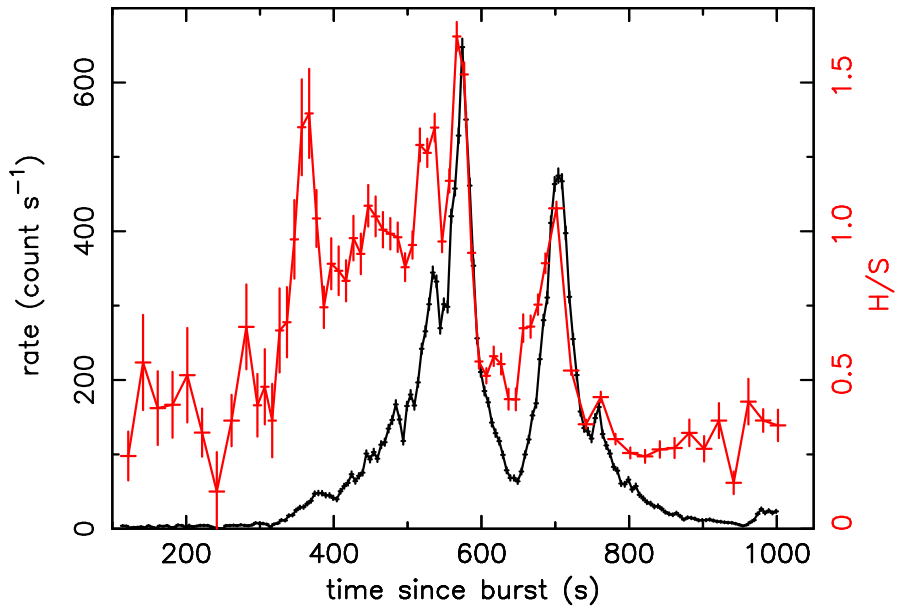}
\end{center}
  \caption{ The observed XRT count-rate light curve for GRB060124
  (black; left-hand scale) and the (2--10 keV / 0.2--2 keV)
  hardness ratio (red; right-hand scale). The spectrum hardens
  during each flaring episode.  }
\label{hard}
\end{figure}

The X-ray flares can show considerable sub-structure. Fig.~\ref{hard}
shows a close-up view of the large flare event in GRB060124. This large
flare shows several episodes of flux increase during which the
spectrum rapidly hardens, followed by a more gradual softening as the
flux declines.  The simplest explanation for the observed spectral
behaviour is the movement of the break energy to higher energies at
the onset of the flare, which then falls to lower energies as the
intensity decreases. This behaviour is consistent with that seen in
gamma-ray flares observed during the prompt phase
\cite{ford95, fen99}.

The rapid rise and fall in flux during these early X-ray flares is
inconsistent with an explanation involving interaction of the external jet
shock with the surrounding medium.  In addition, where large energy output
is seen it is likely due to the central engine being fed matter as late
times due to fragmentation of the progenitor \cite{king05} and/or a clumpy
accretion flow \cite{perna06, proga06}. The timescale over which
large X-ray flares occur is mostly confined to the first hour after the
trigger, consistent with the previously known range in burst duration.

\section{Short bursts}

The discussion above is based mainly on the observed properties of
long GRBs (i.e. T$_{90} > 2$s).  Pre-\swift\ no accurate localizations
had been obtained for short bursts, but now both \swift\ and {\it
HETE-2} have provided data which has allowed for the accurate
localization of several short bursts.  From the first two which were
localized, GRB050509B \cite{ge05} and GRB050709
\cite{fox05}, it was apparent that short bursts are associated
with host galaxies that have less active star formation than long
bursts and that they are in lower density local environments
\cite{ge05, hj05, sod06}.  Follow-up observations have provided
redshifts for most of the localized short bursts and show they have
systematically lower redshifts than the long bursts.  The lower
redshifts and lower fluences imply lower luminosities, although as for
long bursts determining the jet opening angle is problematic
\cite{sod06, bu06}.  The long-burst progenitor is thought to be a collapsar
(section 1) whereas the properties of the short-bursts are consistent
with a neutron star--neutron star or neutron star -- black hole binary
progenitor.  Short-lifetime massive stars (collapsars) are very
unlikely progenitors for the observed short bursts because of both the
lack of recent star formation in the host galaxies and the absence of
a supernova which should be detectable at low redshifts
\cite{hj05, bloom06}.

\begin{figure}
\begin{center}
  \includegraphics[angle=0,width=.65\textheight]{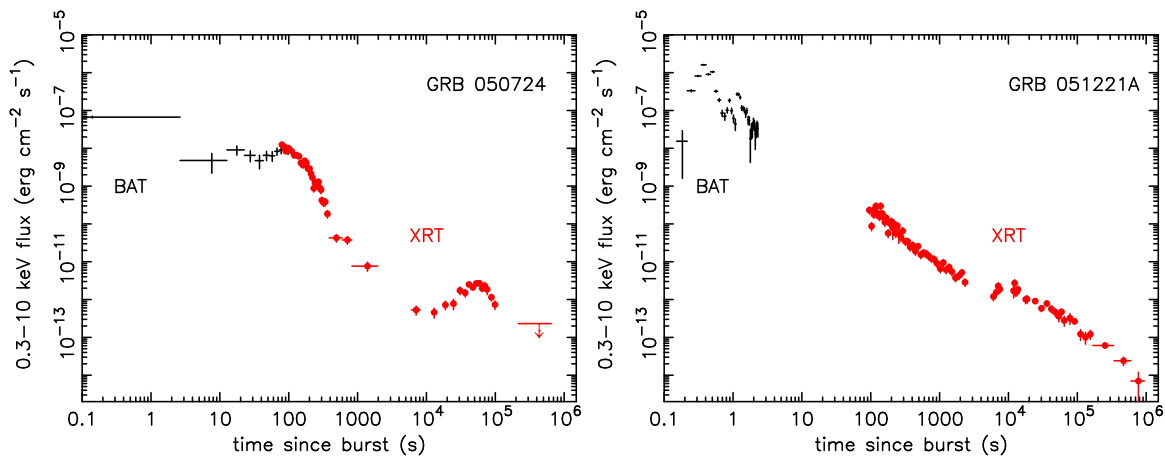}
\end{center}
  \caption{ The BAT+XRT flux light curves for the short bursts
  GRB050724 and GRB051221A. GRB050724 displays a steep decay and
  several flares. GRB051221A is a very bright burst with a gradual
  initial decline, flux $\propto t^{-1}$, a late emission hump from a
  few thousand to $\sim 10^4$s, followed by another gradual decline
  $\propto t^{-1}$.}
\label{short}
\end{figure}

Despite the environmental and likely progenitor differences, the X-ray
light curves of short bursts are very similar to those of long
bursts. In Fig.~\ref{short} the BAT+XRT light curves of two well
studied short bursts display the full range of X-ray phenomena: steep
decay and flares (GRB050724); and gradual decay and late emission hump
(GRB051221A). To \swift\ GRB050724 is technically a long burst,
but it would have appeared as a short burst to the BATSE instrument
\cite{ba05b}. The long duration of X-ray emission for the short bursts
suggests that their central engines can also be fuelled for many
hours, possibly due to the same processes discussed above to explain
X-ray flares.

\section{Conclusions} 

The \swift\ era has truly energized the study of Gamma-ray Bursts. In
its first year of operation \swift\ has provided the first accurate
X-ray localization for a short burst, found the highest redshift and
highest X-ray luminosity source, GRB050904, observed large X-ray
flares which can occur up to an hour or more after the burst and
observed a wide variety of temporal and spectral shapes for GRBs. The
early high-energy emission from most GRBs appears to be dominated by
central engine activity, which may continue low energy output for up to
a day after the burst. This phase, plus X-ray flares, are seen in both
long and short bursts. In a significant minority of GRBs, the early
X-ray emission is consistent with a classical afterglow, where we see
early interaction of the jet with the circum-burst environment. The
wealth of observational phenomena challenge practically all of the
theories as to how GRBs are powered, the nature of the relativistic
jet and the interaction between a GRB and its environment.

\ack The authors gratefully acknowledge funding for \swift\ at the
University of Leicester by PPARC, in the USA by NASA and in Italy by ASI.
We are also very grateful to our colleagues on the \swift\ project for
their help and support. We thank Kim Page for help with the figures.

\end{document}